\documentclass[12pt]{article}

\usepackage[margin=1in]{geometry}
\usepackage{setspace}
\usepackage{amsmath, amssymb, amsthm}
\usepackage{booktabs}
\usepackage{threeparttable}
\usepackage{graphicx}
\usepackage{subcaption}
\usepackage{natbib}
\usepackage{hyperref}
\usepackage{siunitx}

\hypersetup{
  colorlinks=true,
  linkcolor=blue,
  citecolor=blue,
  urlcolor=blue
}

\graphicspath{{figures/}}

\title{\vspace{-0.7cm}
Monitoring Limits in DAO Governance: Capacity Breakpoints and Endogenous Concentration
}

\author{
Guy Tchuente\thanks{Purdue University. Email: \texttt{gtchuent@purdue.edu}.} \\
}

\date{\today}

\begin{document}
\maketitle

\begin{abstract}
Decentralized autonomous organizations (DAOs) are designed to disperse control, yet recent evidence shows that effective governance is often concentrated in a small number of participants.
This note studies one simple mechanism behind that pattern.
Because decentralized governance is monitor-intensive, rising proposal flow may eventually outpace the capacity of broad-based participation.
Using a DAO--quarter panel, I estimate a fixed-effects kink model with DAO and quarter fixed effects and find a statistically significant decline in the marginal responsiveness of active voters once proposal activity crosses an interior threshold.
I then study realized voting concentration using kink specifications with data-driven cutoffs.
Across specifications, decentralization gains do not persist indefinitely once governance workload becomes sufficiently high, and load-based measures show especially clear evidence of a transition toward more concentrated realized control.
The results provide reduced-form evidence consistent with a ``too big to monitor'' mechanism in DAO governance: when proposal flow grows faster than broad participation can keep up, effective control may drift toward a smaller set of highly active participants.
\end{abstract}

\vspace{0.2cm}
\newpage
\section{Introduction}
Decentralized autonomous organizations (DAOs) are often presented as a governance technology capable of dispersing authority through member voting rather than centralized managerial control.
A growing literature documents both the rapid expansion of DAO activity and the institutional diversity of on-chain governance, including variation in voting rules, delegation mechanisms, quorum requirements, and implementation architectures \citep{han2025reviewdao,nassif2024daos,bellavitis2023rise,sharma2024future}.
Yet decentralization is not costless.
For collective decision-making to remain genuinely broad-based, participants must monitor governance activity: they must read proposals, assess tradeoffs, follow discussion, and cast informed votes.
As proposal flow rises, this monitoring task may become increasingly burdensome, creating scope for participation fatigue, reliance on delegates, and the effective concentration of influence even when formal voting rights remain widely distributed \citep{ammons2025idealsinstitutions,zhang2025governancedefi,bongaerts2025votedelegation}.

This note is part of a broader research program on the limits of decentralized monitoring.
In earlier work, I develop and test a ``too big to monitor'' framework showing that decentralized monitoring systems may function well at modest scale but can break down once the volume of entities or decisions to be monitored exceeds the capacity of dispersed monitors \citep{tchuente2025too,tchuente2026scale}.
The first papers in that series study more traditional regulatory and service-delivery environments.
The present note extends the same logic to DAOs, a setting that is institutionally very different and therefore especially revealing.
In DAOs, monitoring is voluntary, governance is token-mediated, participation may be pseudonymous, and authority is designed to be decentralized from the outset.
If a too-big-to-monitor mechanism arises even in this environment, it suggests that capacity limits are a more general constraint on decentralized governance than the earlier applications alone might imply.

The note also speaks directly to recent evidence on concentration in DAO governance.
\citet{appel2023control} show that many DAOs are effectively controlled by a very small number of entities, while \citet{appel2026daogovernance} emphasize that DAO governance outcomes depend critically on institutional design rather than on decentralization alone.
This paper provides a simple empirical micro-foundation for those findings.
The core idea is that scale in a DAO should be understood not only as the number of proposals submitted, but as the monitoring burden those proposals impose on the voting community.
If monitoring capacity is limited, active participation may initially rise with proposal volume but eventually fail to keep pace.
Beyond that point, additional governance activity no longer attracts commensurate broad participation, and the effective conduct of governance may shift toward a smaller set of highly active participants.
In this sense, the relevant consequence of scale is not merely lower turnout, but the endogenous reconcentration of influence within a formally decentralized system.

Empirically, I construct a DAO--quarter panel from proposal and voting activity and study whether decentralized participation exhibits a measurable capacity threshold.
Rather than treating concentration as a static property of token ownership, I ask a dynamic question:
does decentralized participation cease to scale beyond a certain governance workload, and do concentration measures shift systematically in the same region?
This framing complements the earlier papers in the series by moving from public and regulatory monitoring environments to decentralized digital governance, and it complements the DAO literature by focusing not only on whether control is concentrated, but also on one process through which concentration may emerge endogenously.

\paragraph{Contributions.}
This note makes three contributions.
First, it extends the broader too-big-to-monitor research agenda to decentralized digital organizations and provides an empirical test of monitoring-capacity limits in a setting where authority is already formally decentralized.
Second, it estimates a \emph{capacity cutoff} in the relationship between proposal volume and active voter participation using a fixed-effects kink specification with a data-driven breakpoint, and then studies whether concentration outcomes---including the Herfindahl--Hirschman Index and top-3 voting control---shift in the same region, both as functions of proposal scale and of monitoring-load measures such as proposals per voter.
Third, it offers a compact empirical micro-foundation for the concentration patterns documented by \citet{appel2023control,appel2026daogovernance}: when governance workload grows faster than broad participation capacity, effective control may drift toward a smaller set of highly active participants.
More broadly, the note provides a transparent empirical pipeline that can be extended to heterogeneity in delegation, quorum rules, and DAO design, linking naturally to recent work on governance architecture and value creation in DAOs \citep{bellavitis2025votinggovernance,zhang2025governancedefi}.

\section{Data and Measurement}
The analysis uses a DAO--quarter panel constructed from proposal- and vote-level governance records.
Each observation corresponds to DAO \(i\) in calendar quarter \(t\).
The empirical objective is to relate three objects measured within the same DAO--quarter: the scale of governance activity, the breadth of realized participation, and the concentration of effective voting control.
This measurement strategy follows the spirit of \citet{appel2023control}, who study DAO governance using proposal and vote records and show that a small number of entities often control most decisions.
It also complements \citet{appel2026daogovernance}, who emphasize that DAO governance outcomes depend on institutional design and broad participation rather than on decentralization in name alone.

\subsection{Variable construction}
The panel is built by aggregating proposal and voting activity to the DAO--quarter level.
Let \(i\) index DAOs and \(t\) quarters.

\paragraph{Governance scale.}
Let \(P_{it}\) denote the number of governance proposals brought to a vote in DAO \(i\) during quarter \(t\).
This is the paper's baseline measure of governance workload.
The main running variable is
\[
x_{it}=\ln(1+P_{it}),
\]
which allows proportional comparisons across DAOs while preserving observations with small proposal counts.

\paragraph{Participation capacity.}
Let \(V_{it}\) denote the number of \emph{active voters} in DAO \(i\) during quarter \(t\), defined as the number of distinct voting entities (or, when entity-level aggregation is unavailable, distinct voting addresses) that cast at least one vote on at least one proposal during that quarter.
In the note, \(V_{it}\) is interpreted as a revealed measure of monitoring and participation capacity: it captures how many participants actually engage in governance when proposals arrive.
The main outcome in the capacity regressions is
\[
y_{it}=\ln(V_{it}).
\]

\paragraph{Monitoring load.}
A central premise of the note is that governance scale matters not only in levels but also relative to the breadth of active participation.
I therefore define the DAO--quarter monitoring-load measure as proposals per active voter:
\[
\text{load}_{it} \equiv \frac{P_{it}}{V_{it}},
\qquad
\ell_{it}\equiv \ln(1+\text{load}_{it}).
\]
This variable is not used in the baseline capacity regression, which focuses on how active participation scales with proposal volume, but it is used in the concentration analysis as a compact measure of governance burden per active participant.

For robustness, I also use \(N_{it}\), the number of voters recorded in DAO \(i\) during quarter \(t\), to construct an alternative workload measure based on proposals per voter rather than proposals per active voter.

\paragraph{Governance concentration.}
The note studies two outcome measures of concentration, both constructed from realized voting outcomes within a DAO--quarter rather than from token holdings alone.
This choice is deliberate: the question is who effectively participates in and controls decisions, not simply who holds tokens.

Let \(s_{jit}\) denote entity \(j\)'s share of realized voting power cast in DAO \(i\) during quarter \(t\), aggregated across all proposals in that quarter.
Using these within-quarter vote shares, I construct:

\begin{enumerate}
    \item the \emph{Herfindahl--Hirschman Index}
    \[
    HHI_{it}=\sum_j s_{jit}^2,
    \]
    which increases as realized voting influence becomes more concentrated; and

    \item the \emph{Top-3 control share}
    \[
    Top3_{it}=\sum_{j\in \mathcal{T}_{it}^{(3)}} s_{jit},
    \]
    where \(\mathcal{T}_{it}^{(3)}\) denotes the three entities with the largest realized voting shares in DAO \(i\), quarter \(t\).
\end{enumerate}

These measures are closely aligned with the control-oriented perspective in \citet{appel2023control}, who emphasize that DAO decisions are often effectively determined by a very small number of participants.
In the present paper, the contribution is dynamic: rather than treating concentration as a fixed feature of DAO governance, I ask whether concentration changes systematically once governance workload becomes sufficiently high relative to broad participation.

\subsection{Sample and estimation samples}

The analysis focuses on the post-2020 period. I distinguish between the full post-2020 sample and narrower nested estimation samples required by specific specifications. The full sample contains all DAO-quarter observations that survive the baseline cleaning rules. The capacity sample additionally requires nonmissing proposals and active voters so that \(\ln(1+P_{it})\) and \(\ln(V_{it})\) are observed. The harmonized concentration sample additionally requires nonmissing monitoring-load and concentration measures. In the current working panel, the first observed quarter is 2020q2. Table~\ref{tab:desc_dao_post2020} reports sample coverage and descriptive statistics for these nested samples.

I use three nested samples.

\begin{enumerate}
    \item The \emph{full post-2020 sample} contains all DAO--quarters satisfying the baseline data-cleaning rules.

    \item The \emph{capacity sample} requires no missing \(P_{it}\) and \(V_{it}\), so that both
    \(\ln(1+P_{it})\) and \(\ln(V_{it})\) are observed.

    \item The \emph{harmonized concentration sample} additionally requires nonmissing
    \(\ell_{it}=\ln(1+P_{it}/V_{it})\), \(HHI_{it}\), and \(Top3_{it}\).
    This harmonized restriction ensures that comparisons across the concentration specifications are not driven by outcome-specific missingness.
\end{enumerate}

Table~\ref{tab:desc_dao_post2020} reports sample coverage and descriptive statistics.
The table shows that the sample contains 686 DAO--quarters from 136 DAOs over 10 calendar quarters, with estimating samples of 680 DAO--quarters for the capacity analysis and 679 DAO--quarters for the harmonized concentration analysis.
Proposal activity is highly skewed, active participation varies substantially across DAOs and over time, and both \(HHI\) and the Top-3 control share indicate sizable dispersion in realized voting concentration.
These patterns are consistent with the broader DAO evidence in \citet{appel2023control} that formal decentralization often coexists with substantial concentration of effective control.

\begin{table}[!htbp]\centering
\caption{Sample overview and descriptive statistics (POST-2020)}
\label{tab:desc_dao_post2020}
\resizebox{\textwidth}{!}{
\begin{tabular}{lcccccccccccc}
\toprule
 & \multicolumn{4}{c}{Full sample} & \multicolumn{4}{c}{Capacity sample} & \multicolumn{4}{c}{Harmonized concentration sample} \\
\cmidrule(lr){2-5}\cmidrule(lr){6-9}\cmidrule(lr){10-13}
Variable & N & Mean & SD & Median & N & Mean & SD & Median & N & Mean & SD & Median \\
\midrule
Proposals &       680 &    15.384 &    32.542 &     7.000 &       680 &    15.384 &    32.542 &     7.000 &       679 &    15.405 &    32.561 &     7.000 \\
Active voters &       686 &   642.327 &  1851.843 &   123.500 &       680 &   647.107 &  1859.264 &   128.500 &       679 &   646.788 &  1860.616 &   128.000 \\
$\ln(1+\text{proposals})$ &       680 &     2.127 &     1.034 &     2.079 &       680 &     2.127 &     1.034 &     2.079 &       679 &     2.129 &     1.033 &     2.079 \\
$\ln(\text{active voters})$ &       686 &     4.851 &     1.841 &     4.816 &       680 &     4.860 &     1.843 &     4.856 & . & . & . & . \\
Proposals / active voters & . & . & . & . & . & . & . & . &       679 &     0.230 &     0.813 &     0.055 \\
$\ln(1+\text{proposals}/\text{active voters})$ & . & . & . & . & . & . & . & . &       679 &     0.139 &     0.283 &     0.054 \\
HHI & . & . & . & . & . & . & . & . &       679 &     0.300 &     0.218 &     0.236 \\
Top-3 control share & . & . & . & . & . & . & . & . &       679 &     0.695 &     0.376 &     0.889 \\
\midrule
\multicolumn{13}{l}{\textit{Sample overview}} \\
DAO-quarters & \multicolumn{4}{c}{686} & \multicolumn{4}{c}{680} & \multicolumn{4}{c}{679} \\
DAOs & \multicolumn{4}{c}{136} & \multicolumn{4}{c}{135} & \multicolumn{4}{c}{135} \\
Calendar quarters & \multicolumn{4}{c}{10} & \multicolumn{4}{c}{10} & \multicolumn{4}{c}{10} \\
First quarter & \multicolumn{4}{c}{2020q2} & \multicolumn{4}{c}{2020q2} & \multicolumn{4}{c}{2020q2} \\
Last quarter & \multicolumn{4}{c}{2022q3} & \multicolumn{4}{c}{2022q3} & \multicolumn{4}{c}{2022q3} \\
\bottomrule
\end{tabular}
}
\begin{minipage}{0.97\textwidth}
\footnotesize \textit{Notes:} The full sample contains all DAO-quarter observations in the post-2020 working panel that survive the baseline cleaning rules. Reported \(N\) may vary across rows within a sample column because descriptive statistics are calculated using nonmissing observations for each variable. The capacity sample additionally requires nonmissing \(\ln(\text{active voters})\) and \(\ln(1+\text{proposals})\). The harmonized concentration sample additionally requires nonmissing monitoring-load measures, HHI, and Top-3 control share.
\end{minipage}
\end{table}

\section{Empirical Strategy}

This note studies whether broad participation in DAO governance exhibits a measurable capacity limit and whether realized voting concentration changes in the same range of governance workload.
The empirical design is intentionally parsimonious.
All specifications are estimated on DAO--quarter panels with DAO fixed effects and quarter fixed effects, and standard errors are clustered at the DAO level.
The results should therefore be interpreted as reduced-form evidence based on within-DAO changes over time rather than as fully causal estimates.

\subsection{Capacity kink model}
To estimate when decentralized participation becomes capacity constrained, I fit a fixed-effects kink regression of active participation on proposal workload:
\begin{equation}
\label{eq:kink_capacity}
y_{it}=\alpha_i+\gamma_t+\beta_1 x_{it}+\beta_2 (x_{it}-c)_+ + \varepsilon_{it},
\end{equation}
where
\[
x_{it}=\ln(1+P_{it}), \qquad y_{it}=\ln(V_{it}),
\]
\(P_{it}\) is the number of proposals in DAO \(i\) during quarter \(t\), \(V_{it}\) is the number of active voters, \(\alpha_i\) are DAO fixed effects, \(\gamma_t\) are quarter fixed effects, and
\[
(x_{it}-c)_+ \equiv \max\{x_{it}-c,0\}
\]
is the kink term.

In this specification, \(\beta_1\) is the slope below the cutoff and \(\beta_1+\beta_2\) is the slope above the cutoff.
A negative \(\beta_2\) indicates that active participation becomes less responsive to additional proposal flow once workload crosses the threshold \(c\), which is consistent with a monitoring-capacity or attention-capacity constraint.

\subsection{Data-driven breakpoint selection}
I select the breakpoint \(c\) by grid search over candidate values spanning the interior of the running-variable distribution, specifically the 10th--90th percentiles of \(x_{it}\).
For each candidate cutoff, I estimate \eqref{eq:kink_capacity} and compute the residual sum of squares.
The estimated cutoff is the value that minimizes this criterion.

This procedure is descriptive rather than structural.
It is designed to locate the point at which the relationship between proposal workload and active participation changes most sharply in the data.
To assess sensitivity to sampling variation, the appendix reports bootstrap evidence based on cluster resampling at the DAO level.

\subsection{Concentration transitions with free cutoffs}
The second part of the analysis asks whether realized voting concentration also exhibits a transition as governance workload rises.
Rather than imposing the participation cutoff on concentration outcomes, the baseline concentration specifications allow each outcome to choose its own breakpoint.

For a generic concentration outcome \(z_{it}\), I estimate
\begin{equation}
\label{eq:kink_concentration}
z_{it}=\alpha_i+\gamma_t+\delta_1 r_{it}+\delta_2 (r_{it}-c_z)_+ + u_{it},
\end{equation}
where \(z_{it}\) is either the Herfindahl--Hirschman Index \(HHI_{it}\) or the Top-3 control share \(Top3_{it}\), \(r_{it}\) is the relevant running variable, and \(c_z\) is selected separately for each outcome by the same RSS-minimizing grid-search procedure.

I use two running variables.

First, I use proposal scale directly:
\[
r_{it}=\ln(1+P_{it}).
\]
These specifications ask whether concentration changes once raw proposal volume becomes sufficiently large.

Second, I use monitoring load per active participant:
\[
r_{it}=\ell_{it}=\ln\!\left(1+\frac{P_{it}}{V_{it}}\right).
\]
These specifications ask whether concentration changes when governance burden rises relative to the breadth of realized participation.

Allowing \(HHI\) and Top-3 to choose their own cutoffs is important in this setting.
The purpose of the note is not to impose a common threshold across all outcomes, but to examine whether different indicators of effective control display a similar regime-change pattern as governance workload increases.

\subsection*{Robustness to alternative load definitions}
Because the active-load measure \(P_{it}/V_{it}\) uses realized participation in the denominator, I also report robustness checks using an alternative load measure based on the broader voting base:
\[
\ell^{NV}_{it}=\ln\!\left(1+\frac{P_{it}}{N_{it}}\right),
\]
where \(N_{it}\) is the number of voters recorded in DAO \(i\) during quarter \(t\).
I then re-estimate the concentration kink regressions for \(HHI_{it}\) and \(Top3_{it}\) using \(\ell^{NV}_{it}\) as the running variable and allowing each regression to choose its own cutoff.
This robustness exercise helps show that the main concentration patterns are not driven solely by the use of active voters in the denominator.

\subsection*{Interpretation}
The identifying variation in all specifications comes from within-DAO changes in proposal workload and monitoring load over time, after absorbing DAO fixed effects and common quarter shocks.
This design does not imply that proposal flow is exogenous.
Periods with unusually many proposals may also coincide with disputes, treasury events, strategic mobilization, or other DAO-specific shocks.
For that reason, I interpret the estimates as disciplined descriptive evidence consistent with a too-big-to-monitor mechanism, not as definitive proof that proposal volume alone causes lower participation or greater concentration.
The purpose of the empirical strategy is narrower: to document whether governance workload, participation saturation, and realized concentration move together in a way that is consistent with a monitoring-capacity interpretation.
\section{Results}

This section proceeds in three steps.
I first estimate the participation-capacity breakpoint.
I then examine whether realized concentration changes as governance activity rises, using both proposal scale and proposals per active voter.
Finally, I assess robustness to an alternative monitoring-load definition and report bootstrap evidence on cutoff uncertainty.
Throughout, the results are interpreted as reduced-form evidence on within-DAO regime changes rather than as fully causal estimates.

\subsection{Capacity breakpoint: participation stops scaling proportionally with proposal volume}

Table~\ref{tab:capacity_kink} reports the baseline capacity specification, which relates
\(\ln(V_{it})\) to \(\ln(1+P_{it})\) with DAO and quarter fixed effects.
The breakpoint is chosen by grid search to minimize the residual sum of squares in the kink specification.
The estimated cutoff is \(\hat c_{\mathrm{cap}}=2.3441\) in \(\ln(1+\text{proposals})\), which corresponds to approximately
\(e^{2.3441}-1 \approx 9.4\) proposals in a DAO-quarter.

The main result is a statistically significant decline in the marginal responsiveness of active participation once proposal flow crosses this threshold.
Below the cutoff, the estimated slope is \(1.104\); above the cutoff it falls to \(0.601\).
Thus, proposal activity continues to attract participation beyond the threshold, but at a substantially lower marginal rate.
This is the core empirical signature of the note's monitoring-capacity mechanism:
broad participation rises with governance workload at low levels of activity, but eventually fails to keep pace.

\begin{figure}[!htbp]
\centering
\begin{subfigure}{0.49\textwidth}
  \centering
  \includegraphics[width=\textwidth]{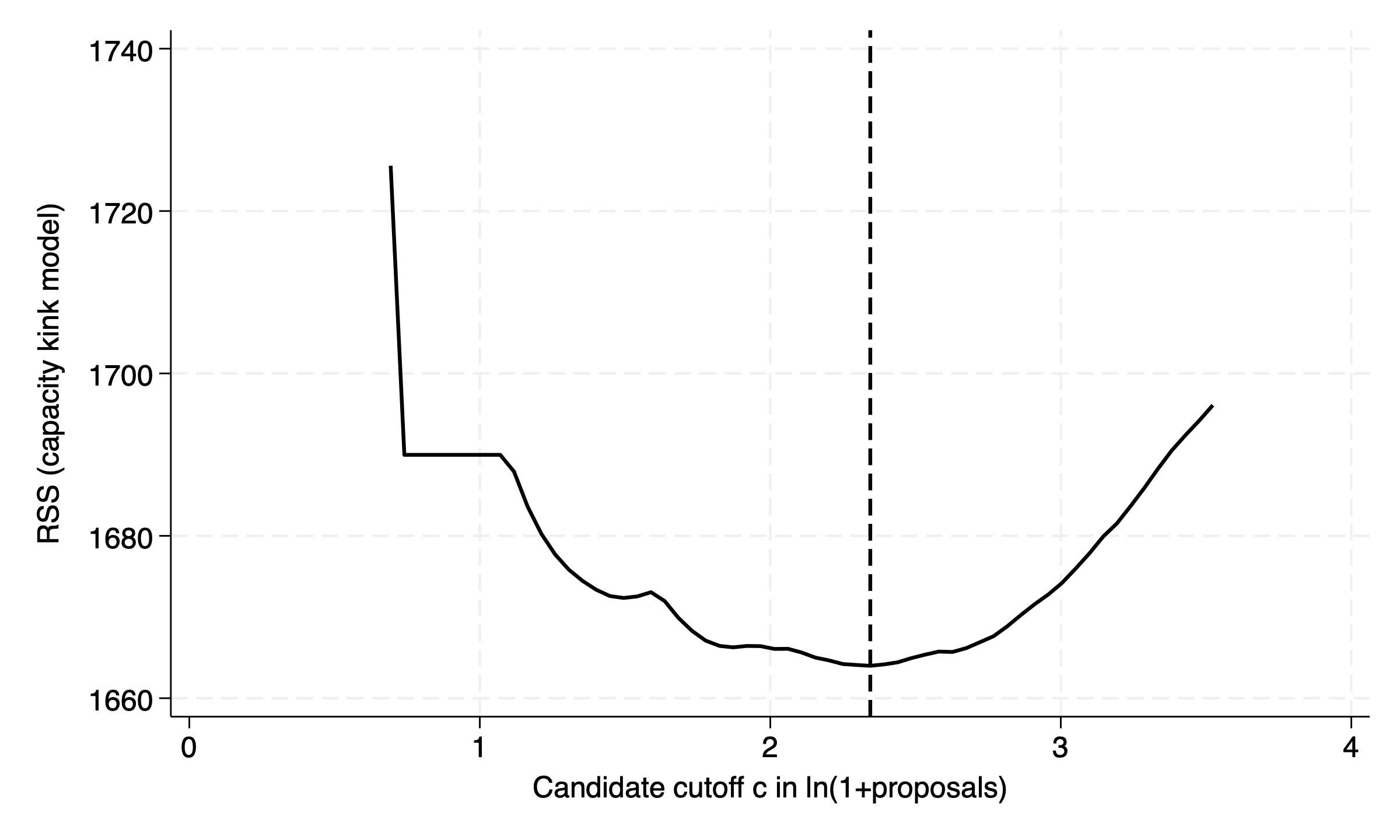}
  \caption{RSS over candidate cutoffs}
\end{subfigure}
\hfill
\begin{subfigure}{0.49\textwidth}
  \centering
  \includegraphics[width=\textwidth]{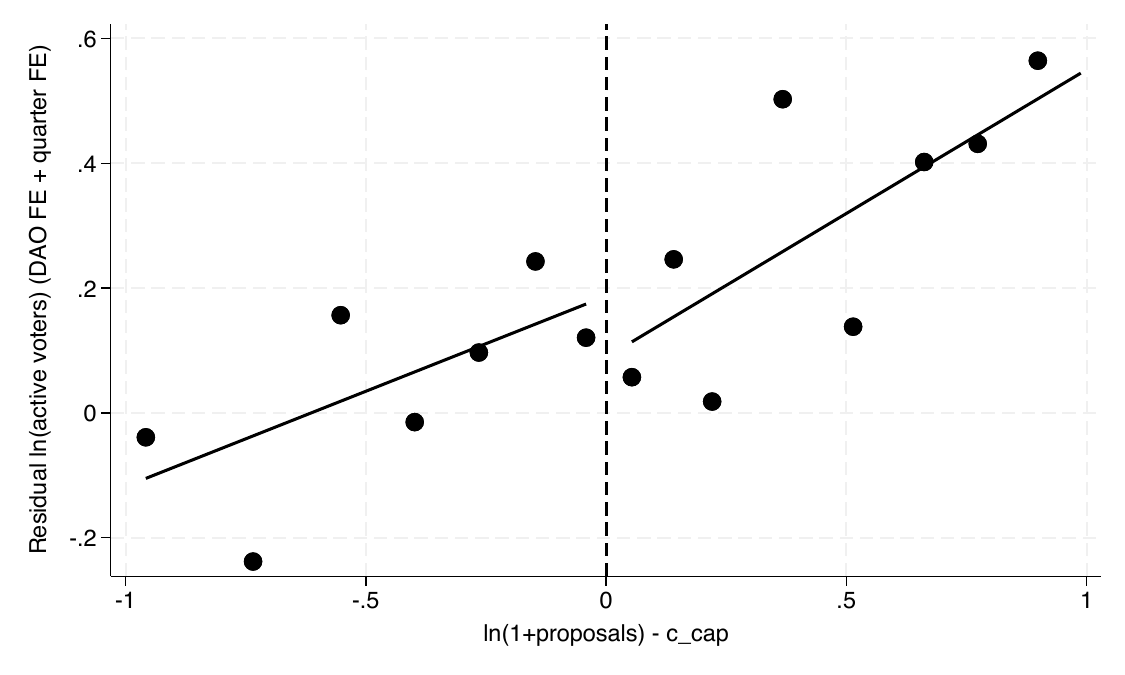}
  \caption{Residualized participation around \(\hat c_{\mathrm{cap}}\)}
\end{subfigure}
\caption{Participation capacity breakpoint}
\label{fig:capacity_breakpoint}
\begin{minipage}{0.92\textwidth}
\footnotesize \textit{Notes:} Panel~(a) plots the residual sum of squares from the fixed-effects kink regression for candidate cutoffs in \(\ln(1+\text{proposals})\). Panel~(b) shows a binned residual plot of \(\ln(\text{active voters})\) after partialling out DAO and quarter fixed effects, centered at the estimated breakpoint.
\end{minipage}
\end{figure}

Figure~\ref{fig:capacity_breakpoint} reinforces this interpretation.
Panel~(a) plots the RSS objective over candidate cutoffs and shows a well-defined interior minimum.
Panel~(b) provides a binned residual plot around the estimated cutoff and makes the change in slope visually transparent.

\begin{table}[!htbp]\centering
\caption{Capacity breakpoint: participation stops scaling proportionally with proposal volume}
\label{tab:capacity_kink}
\begin{tabular}{lcc}
\toprule
 & (1) Linear & (2) Kink \\
 & $\ln(\text{active voters})$ & $\ln(\text{active voters})$ \\
\midrule
$\ln(1+\text{proposals})$ & 0.882*** & 1.104*** \\
 & (0.089) & (0.113) \\
$(\ln(1+\text{proposals})-\hat c_{\text{cap}})_{+}$ &  & -0.503** \\
 &  & (0.198) \\
\midrule
Observations & 680 & 680 \\
Estimated cutoff $\hat c_{\text{cap}}$ &  & 2.3441 \\
Slope below cutoff &  & 1.104 \\
Slope above cutoff &  & 0.601 \\
$p$-value (kink) &  & 0.0123 \\
\bottomrule
\end{tabular}
\begin{minipage}{0.92\textwidth}
\footnotesize \textit{Notes:} The dependent variable is $\ln(\text{active voters})$. All specifications include DAO fixed effects and quarter fixed effects. Standard errors, in parentheses, are clustered at the DAO level. The breakpoint is selected by RSS-minimizing grid search. Sample: capacity estimation sample from the post-2020 working panel. Significance: * $p<0.10$, ** $p<0.05$, *** $p<0.01$.
\end{minipage}
\end{table}

\subsection{Governance concentration under monitoring load and proposal scale}

I next turn from participation to realized concentration.
I begin with a more mechanism-oriented measure of governance burden,
\(\ln(1+P_{it}/V_{it})\), which scales proposal flow by active participation.
This specification is especially relevant for the paper's interpretation, because it more directly captures monitoring load per participant.

Table~\ref{tab:conc_load} shows that the concentration--load relationship is strongly nonlinear.
For \(HHI\), the estimated slope is \(0.978\) below the cutoff and \(0.200\) above it.
For the Top-3 control share, the corresponding slopes are \(5.091\) and \(0.193\).
In both cases, the kink is highly statistically significant.
The pattern is therefore not a reversal but a sharp flattening: as monitoring load rises, concentration increases steeply at low levels of burden and then continues to rise much more slowly beyond the estimated threshold.

\begin{figure}[!htbp]
\centering
\includegraphics[width=0.88\textwidth]{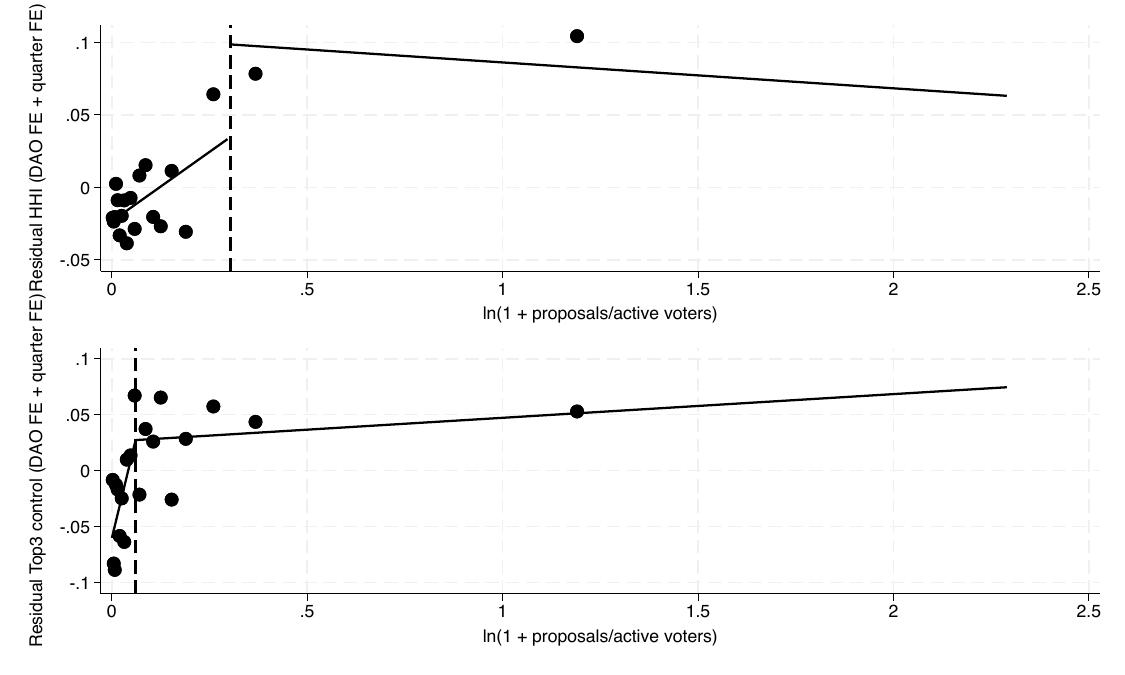}
\caption{Governance concentration and active monitoring load}
\label{fig:conc_load_transition}
\begin{minipage}{0.92\textwidth}
\footnotesize \textit{Notes:} The figure plots binned residual relationships between concentration outcomes and \(\ln(1+\text{proposals}/\text{active voters})\) after partialling out DAO and quarter fixed effects. Each panel is centered at the outcome-specific cutoff selected by the corresponding kink regression. The visual pattern is one of steep increase at low monitoring load followed by substantial flattening beyond the estimated threshold.
\end{minipage}
\end{figure}

Figure~\ref{fig:conc_load_transition} provides a visual counterpart.
The binned residual plots show a steep positive relationship at low levels of active monitoring load and a much flatter relationship after the outcome-specific cutoff.
This is consistent with a saturation mechanism in which increases in governance burden initially coincide with more concentrated realized control, but the marginal effect weakens once the DAO is already operating under high monitoring strain.

\begin{table}[!htbp]\centering
\caption{Governance concentration and active monitoring load}
\label{tab:conc_load}
\resizebox{\textwidth}{!}{
\begin{tabular}{lcccc}
\toprule
 & \multicolumn{2}{c}{HHI} & \multicolumn{2}{c}{Top-3 control share} \\
\cmidrule(lr){2-3}\cmidrule(lr){4-5}
 & (1) Linear & (2) Kink & (3) Linear & (4) Kink \\
\midrule
$\ln(1+\text{proposals}/\text{active voters})$ & 0.371*** & 0.978*** & 0.298*** & 5.091*** \\
 & (0.106) & (0.176) & (0.094) & (0.993) \\
$(\ln(1+\text{proposals}/\text{active voters})-\hat c_z)_{+}$ &  & -0.778*** &  & -4.898*** \\
 &  & (0.192) &  & (0.997) \\
\midrule
Observations & 679 & 679 & 679 & 679 \\
Estimated cutoff $\hat c_z$ &  & 0.3037 &  & 0.0608 \\
Slope below cutoff &  & 0.978 &  & 5.091 \\
Slope above cutoff &  & 0.200 &  & 0.193 \\
$p$-value (kink) &  & 0.0001 &  & 0.0000 \\
\bottomrule
\end{tabular}}
\begin{minipage}{0.97\textwidth}
\footnotesize \textit{Notes:} The dependent variable is either the Herfindahl--Hirschman Index (HHI) of realized voting concentration or the Top-3 control share. Monitoring load is defined as $\ln(1+\text{proposals}/\text{active voters})$. All specifications include DAO fixed effects and quarter fixed effects. Standard errors, in parentheses, are clustered at the DAO level. The breakpoint is selected separately for each outcome by RSS-minimizing grid search. Sample: harmonized concentration sample from the post-2020 working panel. Significance: * $p<0.10$, ** $p<0.05$, *** $p<0.01$.
\end{minipage}
\end{table}

I then return to proposal scale itself, using \(\ln(1+P_{it})\) as the running variable and allowing each concentration outcome to choose its own breakpoint.
This free-cutoff design is intentionally descriptive.
Its purpose is not to force concentration to bend exactly where participation bends, but to ask whether concentration also exhibits a regime change as proposal activity rises.

Table~\ref{tab:conc_prop} shows that for \(HHI\), the concentration--scale relationship reverses sign.
At low-to-moderate proposal volume, concentration declines with scale: the estimated slope below the cutoff is \(-0.046\).
Beyond the estimated threshold, however, the relationship turns positive, with an above-cutoff slope of \(0.029\).
The kink is statistically significant (\(p=0.0321\)), which suggests that the apparent decentralization gains from scale do not persist indefinitely.

For the Top-3 control share, the point estimates tell a similar qualitative story.
The slope is negative below the estimated cutoff and positive above it, but the kink is less precisely estimated.
Taken together, these proposal-scale results are consistent with a gradual transition away from broad-based governance as activity becomes more demanding, although the evidence is stronger for \(HHI\) than for Top-3 concentration.

\begin{figure}[!htbp]
\centering
\includegraphics[width=0.88\textwidth]{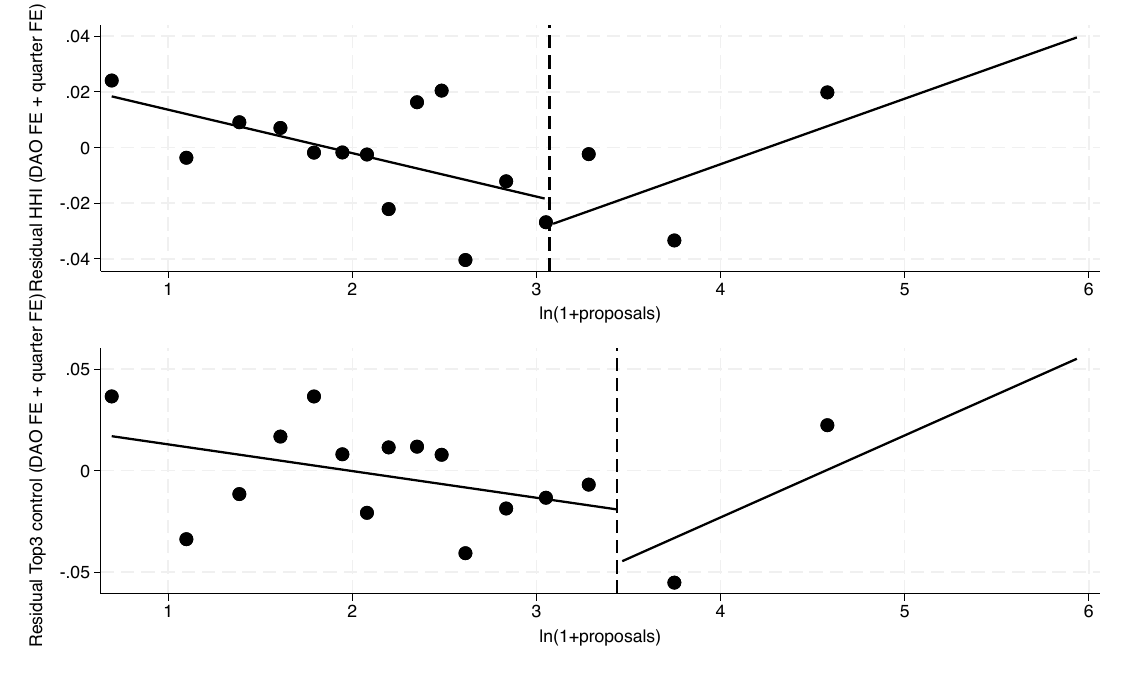}
\caption{Governance concentration and proposal scale}
\label{fig:conc_scale_transition}
\begin{minipage}{0.92\textwidth}
\footnotesize \textit{Notes:} The figure plots binned residual relationships between concentration outcomes and \(\ln(1+\text{proposals})\) after partialling out DAO and quarter fixed effects. Each panel uses the outcome-specific cutoff selected by the corresponding kink regression.
\end{minipage}
\end{figure}

Figure~\ref{fig:conc_scale_transition} provides a visual counterpart.
The binned residual plots again suggest a gradual transition rather than a discrete jump, which is what one would expect from a saturation mechanism in which monitoring capacity becomes progressively strained.

\begin{table}[!htbp]\centering
\caption{Governance concentration and proposal scale}
\label{tab:conc_prop}
\resizebox{\textwidth}{!}{
\begin{tabular}{lcccc}
\toprule
 & \multicolumn{2}{c}{HHI} & \multicolumn{2}{c}{Top-3 control share} \\
\cmidrule(lr){2-3}\cmidrule(lr){4-5}
 & (1) Linear & (2) Kink & (3) Linear & (4) Kink \\
\midrule
$\ln(1+\text{proposals})$ & -0.026 & -0.046** & -0.027 & -0.039 \\
 & (0.016) & (0.021) & (0.021) & (0.025) \\
$(\ln(1+\text{proposals})-\hat c_z)_{+}$ &  & 0.075** &  & 0.097 \\
 &  & (0.035) &  & (0.073) \\
\midrule
Observations & 679 & 679 & 679 & 679 \\
Estimated cutoff $\hat c_z$ &  & 2.8653 &  & 3.5264 \\
Slope below cutoff &  & -0.046 &  & -0.039 \\
Slope above cutoff &  & 0.029 &  & 0.058 \\
$p$-value (kink) &  & 0.0321 &  & 0.1825 \\
\bottomrule
\end{tabular}}
\begin{minipage}{0.97\textwidth}
\footnotesize \textit{Notes:} The dependent variable is either the Herfindahl--Hirschman Index (HHI) of realized voting concentration or the Top-3 control share. The running variable is $\ln(1+\text{proposals})$. All specifications include DAO fixed effects and quarter fixed effects. Standard errors, in parentheses, are clustered at the DAO level. The breakpoint is selected separately for each outcome by RSS-minimizing grid search. Sample: harmonized concentration sample from the post-2020 working panel. Significance: * $p<0.10$, ** $p<0.05$, *** $p<0.01$.
\end{minipage}
\end{table}
\subsection{Robustness: alternative monitoring-load definition}

A natural concern is that monitoring burden may look mechanically high when proposals are scaled by realized participation.
To address this, Table~\ref{tab:robust_load_nv} replaces proposals per active voter with proposals per recorded voters,
\[
\ell^{NV}_{it}=\ln\!\left(1+\frac{P_{it}}{N_{it}}\right),
\]
where \(N_{it}\) is the number of voters recorded in the DAO-quarter.
This is a more conservative denominator, since it scales proposal flow by a broader voting base.

The qualitative pattern is the same.
For both \(HHI\) and the Top-3 control share, concentration rises steeply with this alternative load measure at low levels of burden and then flattens sharply after the estimated cutoff.
For \(HHI\), the slope falls from \(3.635\) below the cutoff to \(0.155\) above it.
For the Top-3 control share, the corresponding slopes are \(2.418\) and \(0.121\).
In both cases, the kink is highly statistically significant (\(p=0.0004\)).
These results indicate that the link between governance burden and concentration is not an artifact of using active voters in the denominator.

\begin{table}[!htbp]\centering
\begin{threeparttable}
\caption{Robustness: governance concentration and alternative monitoring load}
\label{tab:robust_load_nv}
\begin{tabular}{lcccc}
\toprule
& \multicolumn{2}{c}{HHI} & \multicolumn{2}{c}{Top-3 control share} \\
\cmidrule(lr){2-3}\cmidrule(lr){4-5}
& (1) Linear & (2) Kink & (3) Linear & (4) Kink \\
\midrule
$\ln\!\left(1+\frac{\text{proposals}}{\text{number\_of\_voters}}\right)$
& 0.163*** & 3.635*** & 0.160*** & 2.418*** \\
& (0.036) & (0.952) & (0.054) & (0.622) \\
$(\ln(1+\text{proposals}/\text{number\_of\_voters})-\hat c_z)_+$
&  & -3.480*** &  & -2.298*** \\
&  & (0.959) &  & (0.637) \\
\midrule
Observations & 608 & 608 & 608 & 608 \\
Estimated cutoff \(\hat c_z\) &  & 0.0335 &  & 0.0944 \\
Slope below cutoff &  & 3.635 &  & 2.418 \\
Slope above cutoff &  & 0.155 &  & 0.121 \\
\(p\)-value (kink) &  & 0.0004 &  & 0.0004 \\
\bottomrule
\end{tabular}
\begin{tablenotes}[flushleft]\footnotesize
\item \textit{Notes:} Monitoring load is defined as proposals divided by \texttt{number\_of\_voters}. All specifications include DAO fixed effects and quarter fixed effects. Standard errors, in parentheses, are clustered at the DAO level. The cutoff is selected separately for each outcome by RSS-minimizing grid search. Sample: DAO-quarters with nonmissing alternative load, concentration outcomes, and fixed effects.
\item \textit{Significance:} * \(p<0.10\), ** \(p<0.05\), *** \(p<0.01\).
\end{tablenotes}
\end{threeparttable}
\end{table}

\subsection{Bootstrap uncertainty for estimated cutoffs}

Because all breakpoints are selected in a data-driven way, it is useful to assess how sensitive they are to sampling variation.
Table~\ref{tab:bootstrap_cutoffs} reports DAO-level cluster-bootstrap distributions for the estimated cutoffs.
The participation cutoff is reasonably stable: the bootstrap median for \(\hat c_{\mathrm{cap}}\) is \(2.093\), with a 95\% percentile interval of \([1.205,\,3.556]\).
By contrast, the exact location of the concentration cutoffs is estimated less precisely, especially when raw proposal scale is used as the running variable.
The load-based concentration cutoffs are more tightly concentrated, which is consistent with the paper's mechanism: governance burden per participant appears to be more informative than proposal count alone.

I therefore view the bootstrap exercise as supporting the existence of regime change while cautioning against over-interpreting the precise numerical location of any single concentration threshold.

\begin{table}[!htbp]\centering
\begin{threeparttable}
\caption{Bootstrap uncertainty for estimated cutoffs}
\label{tab:bootstrap_cutoffs}
\begin{tabular}{lccccc}
\toprule
Cutoff & Reps. & Mean & P50 & P2.5 & P97.5 \\
\midrule
\(c_{\mathrm{cap}}\): capacity cutoff in \(\ln(1+\text{proposals})\) & 300 & 2.242 & 2.093 & 1.205 & 3.556 \\
\(c_{\mathrm{load,cap}}\): implied load cutoff                      & 300 & 0.020 & 0.012 & 0.003 & 0.057 \\
\(c_{\mathrm{HHI},P}\): HHI cutoff vs.\ proposals                   & 300 & 2.837 & 3.122 & 0.822 & 3.761 \\
\(c_{\mathrm{Top3},P}\): Top-3 cutoff vs.\ proposals                & 300 & 2.075 & 1.757 & 0.741 & 3.725 \\
\(c_{\mathrm{HHI},L}\): HHI cutoff vs.\ active load                 & 300 & 0.292 & 0.290 & 0.035 & 0.421 \\
\(c_{\mathrm{Top3},L}\): Top-3 cutoff vs.\ active load              & 300 & 0.075 & 0.055 & 0.039 & 0.278 \\
\bottomrule
\end{tabular}
\begin{tablenotes}[flushleft]\footnotesize
\item \textit{Notes:} DAO-level cluster bootstrap. In each replication, DAOs are resampled with replacement and the full cutoff-selection procedure is repeated. P50 denotes the bootstrap median. The capacity cutoff is estimated more stably than the concentration cutoffs, and the load-based concentration cutoffs are more tightly concentrated than the proposal-based cutoffs.
\end{tablenotes}
\end{threeparttable}
\end{table}
\section{Conclusion}
This note studies a simple limit of decentralized governance: broad participation may not scale indefinitely with governance workload.
Using DAO--quarter data, I show that active participation rises with proposal activity up to an interior breakpoint and then becomes substantially less responsive.
This pattern is consistent with a monitoring-capacity constraint: as proposal flow increases, the burden of following, evaluating, and voting on governance decisions begins to outpace the capacity of dispersed participants.

I then show that realized voting concentration exhibits related transition patterns.
When governance burden becomes sufficiently high, concentration outcomes cease to evolve in the same way they do at lower levels of activity.
This is especially clear when workload is measured relative to the breadth of participation.

Overall, the results suggest that decentralization gains become harder to sustain once governance activity becomes sufficiently demanding.  I present a simple mechanism linking the concentration facts documented by \citet{appel2023control,appel2026daogovernance} to the scale of governance activity itself: when proposal flow grows faster than broad participation can keep up, effective control may drift toward a smaller set of highly active participants.

The paper is intentionally modest in scope.
The estimates are reduced-form and should be read as descriptive evidence on within-DAO regime changes rather than as definitive causal effects of proposal flow.
Even so, the patterns are informative.
They suggest that formal decentralization does not eliminate organizational capacity constraints; instead, those constraints may reappear as participation saturation and renewed concentration of influence.

More broadly, the note extends the ``too big to monitor'' framework to decentralized digital organizations \citep{tchuente2025too, tchuente2026scale}.
The main implication is not that DAOs inevitably centralize, but that decentralized governance may require institutional responses---such as delegation design, proposal screening, agenda management, or other participation-saving mechanisms---once governance workload grows sufficiently large.

\bibliographystyle{aer}
\bibliography{ref_cent}

@article{han2025reviewdao,
  title   = {A review of DAO governance: Recent literature and emerging trends},
  author  = {Han, Jungsuk and Lee, Jongsub and Li, Tao},
  journal = {Journal of Corporate Finance},
  volume  = {91},
  pages   = {102734},
  year    = {2025},
  doi     = {10.1016/j.jcorpfin.2025.102734}
}

@article{bellavitis2025votinggovernance,
  title   = {Voting governance and value creation in decentralized autonomous organizations (DAOs)},
  author  = {Bellavitis, Cristiano and Momtaz, Paul P.},
  journal = {Journal of Business Venturing Insights},
  volume  = {24},
  pages   = {e00537},
  year    = {2025},
  doi     = {10.1016/j.jbvi.2025.e00537}
}

@misc{bongaerts2025votedelegation,
  title         = {Vote Delegation in {DeFi} Governance},
  author        = {Bongaerts, Dion and De Jong, Frank and Driessen, Joost and Makarov, Igor and Reshchikov, Nikolai},
  year          = {2025},
  eprint        = {2503.11940},
  archivePrefix = {arXiv},
  primaryClass  = {q-fin.GN},
  doi           = {10.48550/arXiv.2503.11940},
  url           = {https://arxiv.org/abs/2503.11940}
}

@article{nassif2024daos,
  title   = {Decentralized Autonomous Organizations: Governance and Implementation},
  author  = {Nassif, Basilio and Savva, Andreas},
  journal = {Applied Sciences},
  year    = {2024},
  volume  = {14},
  number  = {16},
  pages   = {7007},
  doi     = {10.3390/app14167007}
}

@techreport{ammons2025idealsinstitutions,
  title       = {From Ideals to Institutions: Transaction Costs, Risk, and Governance in Decentralized Autonomous Organizations},
  author      = {Ammons, John and Makridis, Christos},
  institution = {SSRN},
  type        = {SSRN Scholarly Paper},
  number      = {5377155},
  year        = {2025},
  month       = aug,
  url         = {https://papers.ssrn.com/abstract=5377155}
}

@techreport{appel2026daogovernance,
  title       = {{DAO} Governance: Decentralized But Not So Disorganized},
  author      = {Appel, Ian and Grennan, Jillian},
  institution = {SSRN},
  type        = {SSRN Scholarly Paper},
  number      = {6065407},
  year        = {2026},
  month       = jan,
  url         = {https://papers.ssrn.com/abstract=6065407}
}

@techreport{zhang2025governancedefi,
  title       = {Governance in Decentralized Finance ({DeFi}): A Comparative Study of On-Chain Voting and Delegation},
  author      = {Zhang, Junzi},
  institution = {National Bureau of Economic Research},
  type        = {Working Paper},
  number      = {33639},
  year        = {2025},
  month       = apr,
  url         = {https://www.nber.org/papers/w33639}
}

@article{bellavitis2023rise,
  title   = {The rise of decentralized autonomous organizations ({DAOs}): a first empirical glimpse},
  author  = {Bellavitis, Cristiano and Fisch, Christian and Momtaz, Paul P.},
  journal = {Venture Capital},
  year    = {2023},
  doi     = {10.1080/13691066.2023.2297702}
}

@article{appel2023control,
  title   = {Control of Decentralized Autonomous Organizations},
  author  = {Appel, Ian and Grennan, Jillian},
  journal = {AEA Papers and Proceedings},
  year    = {2023},
  volume  = {113},
  pages   = {182--185},
  doi     = {10.21428/58320208.ebd76eea}
}

@misc{sharma2024future,
  title         = {Future of Algorithmic Organization: Large-Scale Analysis of Decentralized Autonomous Organizations ({DAOs})},
  author        = {Sharma, Tanusree and Potter, Yujin and Pongmala, Kornrapat and Wang, Henry and Miller, Andrew and Song, Dawn and Wang, Yang},
  year          = {2024},
  eprint        = {2410.13095},
  archivePrefix = {arXiv},
  primaryClass  = {cs.SI},
  doi           = {10.48550/arXiv.2410.13095},
  url           = {https://arxiv.org/abs/2410.13095}
}

@article{tchuente2026scale,
  title={Scale and Capacity Limits in Decentralized FDA Food-Safety Enforcement},
  author={Tchuente, Guy},
  journal={arXiv preprint arXiv:2602.12392},
  year={2026}
}

@article{tchuente2025too,
  title={Too Big to Monitor? Network Scale and the Breakdown of Decentralized Monitoring},
  author={Tchuente, Guy},
  journal={arXiv preprint arXiv:2511.23320},
  year={2025}
}

\appendix
\section{Appendix: Suggested figure to motivate the setting and data source}
\begin{figure}[!htbp]
\centering
\includegraphics[width=0.88\textwidth]{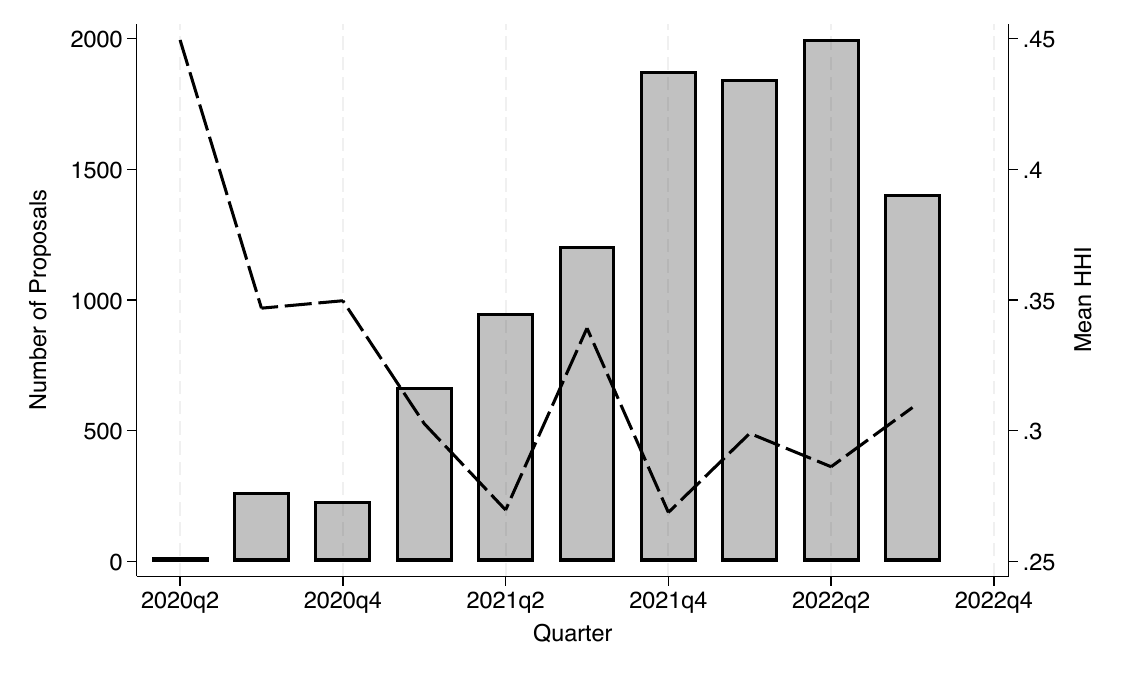}
\caption{Trends in proposals and voting concentration (motivating evidence)}
\label{fig:appel_grennan}
\end{figure}

Figure~\ref{fig:appel_grennan} provides a simple motivating fact in the DAO--quarter panel:
proposal volume rises sharply over time, while average voting concentration remains substantial.
This pattern is consistent with related evidence on concentration in DAO governance from \cite{appel2023control}.

The data used in this paper are constructed from the replication materials associated with \cite{appel2023control}, together with proposal- and vote-level governance files used to build the DAO-quarter panel. From these source files, I construct measures of proposal volume, active voter participation, monitoring load, and realized voting concentration. The final estimation samples are obtained by applying the sample restrictions described in the paper.

\end{document}